\documentclass[aps,showpacs,superscriptaddress,amsmath,amssymb,preprint]{revtex4}
\usepackage{graphicx} 
\usepackage{amsmath}
\usepackage{xcolor}
\usepackage{xcolor}
\usepackage{array}
\usepackage{multirow}  
\usepackage{caption}   
\begin{document}
\title{From first to second minimum: Parity-dependent level densities in $^{240,242}$Pu}

\author{A. Rahmatinejad}\email{a_rahmatinejad@theor.jinr.ru}
\affiliation{Joint Institute for Nuclear Research, Dubna, 141980, Russia}
\author{T. M. Shneidman}
\affiliation{Joint Institute for Nuclear Research, Dubna, 141980, Russia}

\author{N. Jovancevic}
\affiliation{Department of Physics, Faculty of Science, University of Novi Sad,
Trg Dositeja Obradovica 3, Novi Sad, Serbia}

\begin{abstract}
We calculate the parity-dependent level density ratios for $^{240,242}$Pu across a broad range of quadrupole deformations, from the spherical configuration up to the superdeformed region, explicitly including both the ground-state minimum and the second minimum (fission isomer).
The parity-equilibration energy, defined as the excitation energy at which positive- and negative-parity level densities approach equilibrium, is compared between configurations.  A significant reduction is observed near the second minimum, indicating a faster equilibration process in this region.


\end{abstract}
\maketitle

\section{\label{Sec1} Introduction}

Nuclear level density (NLD) is a key quantity in nuclear structure and reaction studies, with wide applications in cross-section calculations and astrophysics. Since Bethe’s introduction of the Fermi Gas Model (FGM) \cite{Bethe1936}, numerous refinements have been developed to account for spin, parity, deformation, and shell effects in the nuclear level density. In particular, the assumption of equal parity distribution at all excitation energies in the Fermi gas model has been shown to break down at low and intermediate energies. Both experimental and theoretical studies confirm that parity asymmetry in nuclear level densities can persist well above the ground state before gradually vanishing as excitation energy increases \cite{Cerf1993, Singhal2011}.

The parity dependence of NLD has been investigated within several microscopic frameworks, including shell-model Monte Carlo calculations \cite{Ozen2007, Alhassid1999} and methods based on quasi-particle excitations combined with BCS occupation numbers \cite{Mocelj2007}. These approaches have highlighted the influence of the underlying single-particle structure, pairing correlations, and collective effects on parity equilibration. Recent studies of transitional and deformed nuclei further demonstrate that parity equilibration is influenced by deformation, and shell closures \cite{Razavi2020, Nuri2021}.

The influence of parity distributions among excited nuclear levels on neutron-capture cross sections in medium-mass nuclei has been explored previously in Refs.~\cite{RauscherPRC1997, Mengoni-NC-1986, Pichon-NPA-1994}, where the analysis was restricted to nuclei at their ground-state deformations. In this work, we extend that study to actinide nuclei, examining how parity equilibration behaves as a function of deformation and shell structure.
In actinide nuclei, large quadrupole deformations lead to multiple minima in the potential-energy surface. In particular, $^{240,242}$Pu exhibit both a ground-state minimum and a superdeformed second minimum (fission isomer). These minima are associated with distinct single-particle structures and shell corrections, and therefore provide an interesting case for studying parity equilibration under different structural conditions.
This study can provide useful input for evaluating the isomer population probabilities against prompt fission probabilities in Pu isotopes, as discussed for example in Ref.~\cite{Goerlach1978}.

The results obtained are conducted within the frame of superfluid model \cite{Bardeen1957,origin1,Dekowski1968} which was previously used for various applications in actinides and superheavy mass regions \cite{Rahmatinejad2021,Azam_ratios_2022,Bezbakh2014}.

\section{Superfluid formalism}
Assuming thermal equilibrium between the neutron and proton subsystems, the nuclear level density is calculated as
\begin{equation} \label{equation13}
\rho(N,Z,U)=\frac{\exp(S)}{{(2\pi)}^{3/2}|D|^{1/2}},
\end{equation}
where $S$ is the entropy of the nucleus and
D is the determinant of the matrix made up of the second derivatives of the logarithm of the partition function $Z$ with respect to inverse temperature $\beta=1/T$ and chemical potentials $\lambda_{\tau}$ of neutrons ($\tau=n$) and protons $(\tau=p)$ \cite{Rahmatinejad2020}.

For each neutron and proton subsystems, the excitation energy $U_{\tau}$ and entropy $S_{\tau}$ of the nucleus is calculated using standard thermal relations:
\begin{equation} \label{equation8}
E_{\tau}=-\frac{\partial \ln Z_{\tau}}{\partial \beta}=\sum_k\left(1-\frac{\varepsilon_k^{\tau}-\lambda_{\tau}}{E^{\tau}_{k}}\tanh\left(\frac{\beta E_{k}^{\tau}}{2}\right)\right)\varepsilon_k^{\tau}-\frac{\Delta_{\tau}^{2}}{G_{\tau}},
\end{equation}
\begin{equation*} \label{equation9a}
U_{\tau}=E_{\tau}-E^{\tau}_{ground},
\end{equation*}
\begin{equation} \label{equation10}
S_{\tau}=\ln Z_{\tau}-\beta \lambda_{\tau} N_{\tau}+\beta E_{\tau}=2 \sum_{k} \ln\left[1+\exp(-\beta E^{\tau}_{k})\right]\\+2\beta\sum_{k}\frac{E^{\tau}_k}{1+\exp(\beta E^{\tau}_{k})},
\end{equation}
where $E^{\tau}_{ground}$ is the ground state energy and $N_{\tau}$ is the number of neutrons or protons.
Neglecting small variations in $\lambda_{\tau}$ with temperature, the heat capacity is obtained as a temperature dependent quantity \cite{RazaviNPA2014,Rahmati2015}:
\begin{equation} \label{equation11}
C_{\tau}=\beta^2 \frac{\partial^2 \ln Z_{\tau}}{\partial \beta^2}=\frac{1}{2}\sum_{k}\cosh^{-2}\left(\frac{\beta E^{\tau}_{k}}{2}\right)\left[\beta^2 E^{\tau2}_{k}-\beta \Delta_{\tau} \frac{d\Delta_{\tau}}{dT}\right].
\end{equation}
The total thermal quantities of the nucleus are obtained as the summation over neutrons and protons:
\begin{eqnarray}
U=U_{n}+U_{p},\nonumber\\
S=S_{n}+S_{p},\nonumber\\
    C=C_{n}+C_{p}.
\end{eqnarray}

In Eqs.~(\ref{equation8}-\ref{equation11}), $E^{\tau}_{k}=\sqrt{(\varepsilon^{\tau}_k-\lambda_{\tau})^2+\Delta_{\tau}^2}$ is the quasi-particle energy
and $\varepsilon_k^{\tau}$ are the single-particle energies obtained using Nilsson shell model Hamiltonian in this work \cite{Nilsson1995}. At each given temperature the pairing gap $\Delta_{\tau}$ and $\lambda_{\tau}$ are obtained by solving the finite-temperature BCS equations:
\begin{equation} \label{equation6}
\frac{2}{G_{\tau}}=\sum_k\frac{1}{E^{\tau}_{k}} \tanh\left(\frac{\beta E^{\tau}_{k}}{2}\right),
\end{equation}
and
\begin{equation} \label{equation7}
N_{\tau}= \frac{1}{\beta}\frac{\partial \ln Z_{\tau}}{\partial \beta}=\sum_k \left(1-\frac{\varepsilon^{\tau}_{k}-\lambda_{\tau}}{E^{\tau}_{k}}\tanh\left(\frac{\beta E^{\tau}_{k}}{2}\right)\right).
\end{equation}
In our study the pairing strength constants $G_{\tau}$ are adjusted to reproduced even-odd mass differences at the ground-state deformation. Further, the same constants are used to calculate pairing gaps for other deformations. The ground-state masses used in these calculations are taken from Ref.~\cite{Moller2016}.

The pairing parameter calculated with the gap equation in the BCS model experiences a sudden decrease at critical temperature $T_{cr}$. However, experimental observations such as S-shaped heat capacity curve obtained from experimental level densities instead of the discontinuity expected from BCS model, show soft changes around the critical temperature for nuclei \cite{Schiller2001,Rahmati2015}.
To avoid the sharp change in pairing that strongly influences the heat capacity, we use the following modified pairing parameter with temperature dependence \cite{Rahmati2015}:
\begin{equation} \label{equation16}
\Delta_{\tau}=\frac{\Delta_0^{\tau}}{1+\exp\left(\frac{T-T_m}{f_{m}}\right)}.
\end{equation}
 Here, $\Delta^{\tau}_{0}$ is the pairing at zero temperature. $T_{m}$ and $f_{m}$ are free parameters that are determined by fitting of numerical calculations in the temperatures larger than 60\% of $T_{cr}$.To ensure a smooth connection between the numerical results and the analytic form of Eq.~\eqref{equation16}, we adopt the following prescription: for $T\leqslant 0.8T_{cr}$ we use the directly calculated $\Delta_{\tau}(T)$, whereas for $T>0.8T_{cr}$ we employ the parametrization of Eq.~\eqref{equation16}.

\section{Parity projection}
In order to obtain parity projected level densities we use the method proposed in Refs.~\cite{Mocelj2007,Alhassid-NPA-1992,Alhassid2000}. The nuclear single-particle levels are divided into groups with the same parity as the whole nucleus at the ground state $\pi^{g}$, and with the parity opposite to the ground state $\pi^{s}$.
In even-even nuclei, as for $^{240,242}$Pu, $\pi^{g}$ is always positive but for odd-A nuclei depending on the parity of the last occupied level by the odd particle, $\pi^{g}$ can be negative or positive. In this work only even-even nuclei are considered ($\pi^{s}=\pi^{-}, \pi^{g}=\pi^{+}$).
Assuming independent and random occupation of single particle states, the probability to find a particle in the group with $\pi^{-}$ is described by the Poisson distribution \cite{Mocelj2007, Alhassid2000}
\begin{equation} \label{eq6}
P(n)=\frac{f^n}{n!}e^{-f},
\end{equation}
where $f$ is the average number of particles in this group.

By summing the occupation numbers of quasi-particle states belonging to the negative-parity group, the mean value of $f_{\tau}$ is estimated as
\begin{equation} \label{eq7}
f_{\tau}=\sum_{k\in\pi^{-}}{\frac{1}{1+\exp(\beta E^{\tau}_{k})}}.
\end{equation}

To construct a negative-parity state of the nucleus, an odd number of particles is required to be excited to the states with negative parity. Therefore, probabilities to find the nuclear system in negative $P^{-}$ and positive $P^{+}$ parity states are given by \cite{Alhassid-NPA-1992}
\begin{eqnarray} \label{a1-0}
P^{-}=\sum_{n}^{odd}\frac{f^n}{n!}e^{-f}=e^{-f} \sinh f,\nonumber \\
P^{+}=\sum_{n}^{even}\frac{f^n}{n!}e^{-f}=e^{-f} \cosh f,
\end{eqnarray}
where $f=f_{n}+f_{p}$.
Dividing the total partition function into partition functions for negative $Z^{-}$ and positive $Z^{+}$ parity states
\begin{equation} \label{a1-1}
    Z=Z^{+}+Z^{-},
\end{equation}
the probabilities $P^{\pm}$ can also be written as
\begin{equation} \label{a1-2}
   P^{\pm}=\frac{Z^{\pm}}{Z}.
\end{equation}
Using Eqs.~(\ref{a1-0}-\ref{a1-2}), we have
\begin{eqnarray}  \label{a50}
  &&  \frac{Z^{-}}{Z^{+}}=\frac{P^{-}}{P^{+}}=\tanh f,\nonumber\\
  &&  \ln(Z^{+})=\ln(Z)-\ln\left(1+\tanh f\right),\nonumber\\
  && \ln(Z^{-})=\ln(Z)-\ln\left(1+\frac{1}{\tanh f}\right).
\end{eqnarray}

The parity projected thermal energies and heat capacities can then be calculated as
\begin{eqnarray}
E^{+}=-\frac{\partial \ln Z^{+}}{\partial \beta}=E+\left[(1-\tanh f)\frac{\partial f}{\partial\beta}\right] \label{a4} \\
E^{-}=-\frac{\partial \ln Z^{-}}{\partial \beta}=E+\left[(1-\coth f)\frac{\partial f}{\partial\beta}\right] \label{a5} \\
C^{+}=\beta^2 \frac{\partial^2 \ln Z^+}{\partial \beta^2}=C-\beta^{2}(1-\tanh(f))\left[(1+\tanh f)\left(\frac{\partial f}{\partial\beta}\right)^{2}-\left(\frac{\partial^{2} f}{\partial\beta^{2}}\right)\right]  \label{a6} \\  \label{a7}
C^{-}=\beta^2 \frac{\partial^2 \ln Z^-}{\partial \beta^2}=C+\beta^{2}(1-\coth(f))\left[(1+\coth f)\left(\frac{\partial f}{\partial\beta}\right)^{2}-\left(\frac{\partial^{2} f}{\partial\beta^{2}}\right)\right].
\end{eqnarray}

Using Eqs.~(\ref{a50}-\ref{a7}), at a given excitation energy $U$, the ratio of the level densities with different parities $R$ is given as \cite{Mocelj2007}
\begin{equation} \label{a03}
R=\frac{\rho^{-}}{\rho^{+}}=\frac{\beta^{-}}{\beta^{+}}\frac{Z^{-}}{Z^{+}}\sqrt{\frac{C^{+}}{C^{-}}}e^{(\beta^{-}-\beta^{+})U},
\end{equation}
where, $\beta^{\pm}$ values are evaluated at each given thermal energy $E=U+E_{ground}$ as
\begin{eqnarray}
E^{+}(\beta^{+})=E^{-}(\beta^{-})=E.
\end{eqnarray}

\section{Results and Discussion}

\begin{figure}[h!]
\begin{center}
\includegraphics[width=0.55\textwidth] {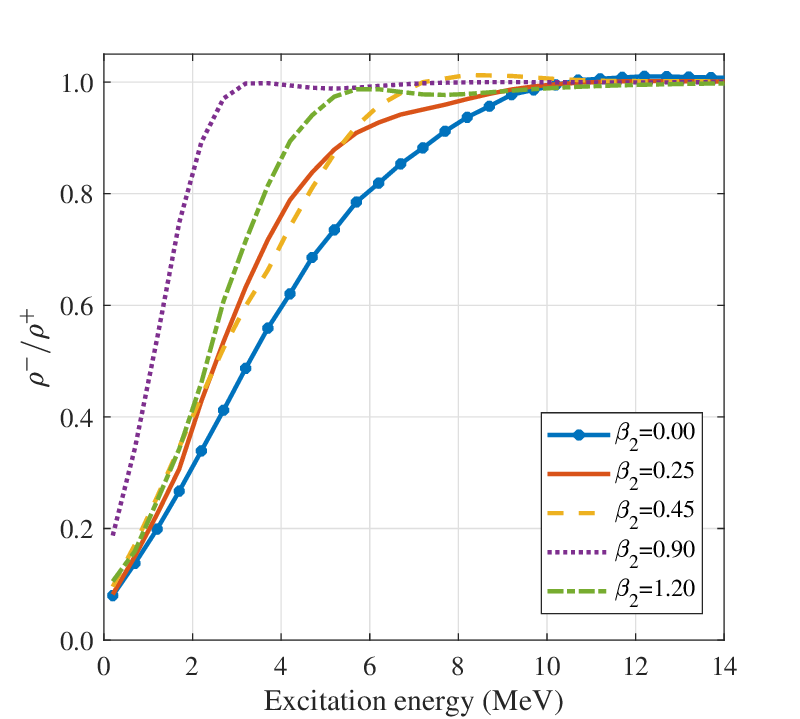}
\end{center}
\caption{Ratios of negative- to positive-parity level densities as a function of excitation energy for $^{240}$Pu at various quadrupole deformations, as indicated in the legend.}
\label{Ru}
\end{figure}

\begin{figure}[h!]
\begin{center}
\includegraphics[width=0.55\textwidth] {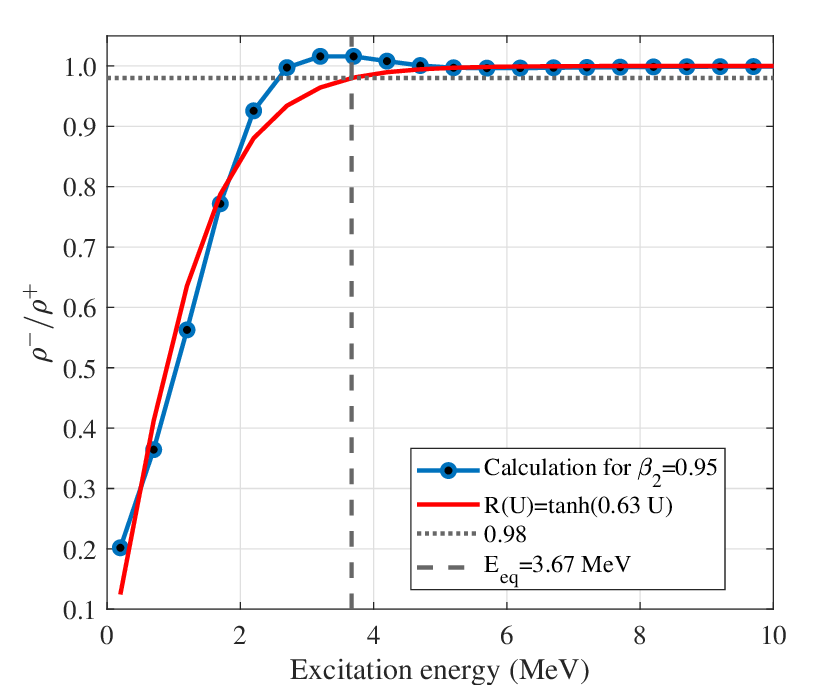}
\end{center}
\caption{The ratio of negative- to positive-parity level densities $R=\rho^{-}/\rho^{+}$ as a function of excitation energy for $^{240}$Pu at quadrupole deformation $\beta_2=0.95$ (blue line with circles). The analytical expression of Eq.~\eqref{Eqend} is shown for comparison (red solid line). The horizontal dotted line marks $R=0.98$, and the vertical dashed line indicates the corresponding excitation energy, defined as the equilibration energy $E_{eq}$.}
\label{RuTan}
\end{figure}

The ratio of the level densities with different parities $R$ is calculated for $^{240,242}$Pu isotopes in the quadrupole deformation range $\beta_2 \in [0,1.25]$ at excitation energies $U = 0.5{-}20$~MeV.
The energy dependence of $R$ for some selected values of $\beta_2$ is shown in Fig.~\ref{Ru} for $^{240}$Pu.
In general, $R$ increases with excitation energy and gradually approaches unity, although the equilibration energy differs for each deformation. In some cases $R$ overshoots unity before returning to it. An example of this behavior is shown in Fig.~\ref{RuTan}.
This overshoot happens when the last occupied orbital is at a shell closure or near a pronounced single-particle gap.
As excitation energy increases in such configurations, a dense group of levels with the parity opposite to the ground state becomes available almost simultaneously, which causes a rapid increase in $R$ and, in certain cases, leads to overshoot $R>1$.

To quantify the rate of parity equilibration at different deformations, the calculated $R$ values are fitted using the function
\begin{equation} \label{Eqend}
    R(U) = \tanh(\gamma U),
\end{equation}
where $\gamma$ is a free parameter characterizing the rate of equilibration.
The equilibration energy $E_{eq}$ is then defined as the excitation energy at which $R(U) = 0.98$.
This threshold is chosen to ensure that $R$ has stabilized near unity, beyond any temporary overshoots where $R > 1$ or significant asymmetries where $R < 1$ (see, for example, Fig.~\ref{RuTan}).

\begin{figure}[h!]
\begin{center}
\includegraphics[width=0.55\textwidth] {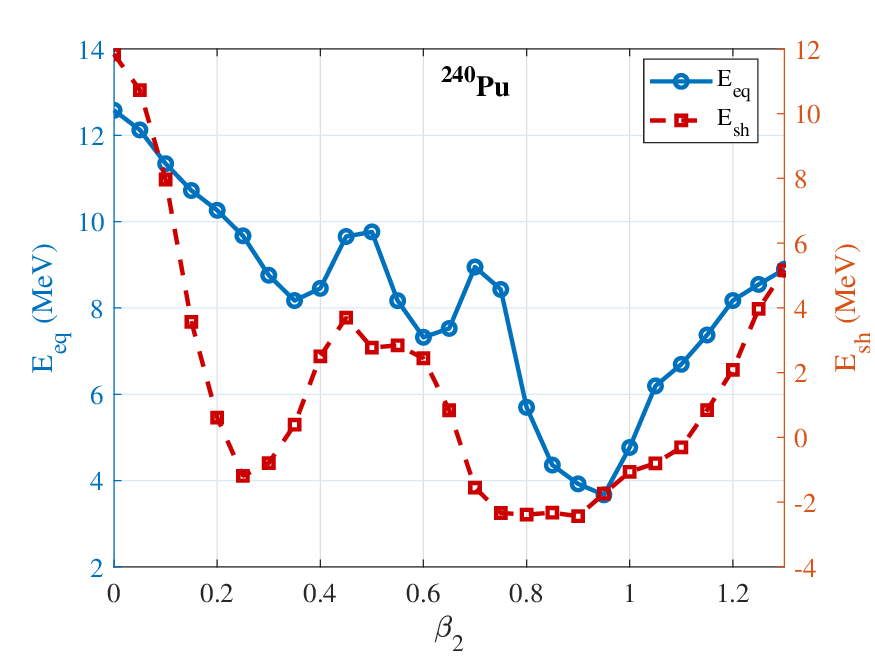}
\includegraphics[width=0.55\textwidth] {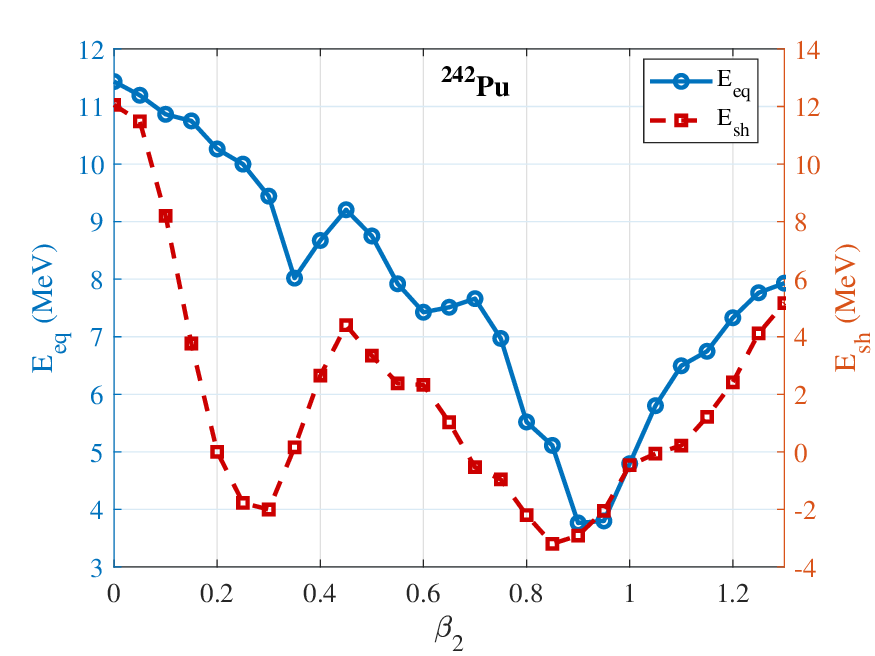}
\end{center}
\caption{Parity equilibration energy $E_{eq}$
 (left vertical axis, blue solid line with circles) and shell correction energy $E_{sh}$
 (right vertical axis, red dashed line with squares) as functions of quadrupole deformation $\beta_{2}$. The upper panel corresponds to $^{240}$Pu and the lower panel to $^{242}$Pu.}
\label{Fig2}
\end{figure}

In Fig.~\ref{Fig2}, the values of $E_{eq}(\beta)$ are shown with the blue line. In general, the calculated equilibration energies for the heavy nuclei $^{240,242}$Pu are systematically smaller than those obtained for lighter systems such as the Mo and Dy isotopes at their ground-state deformations \cite{Razavi2020,Nuri2021}. This trend is consistent with the mass-number dependence of parity asymmetry in nuclear level densities reported in Ref.~\cite{Singhal2011}, and supports the expectation that heavier nuclei equilibrate parities faster due to the higher single-particle level density.

The obtained $E_{eq}(\beta_{2})$ decreases overall with increasing $\beta_{2}$, while several well-pronounced minima are visible. In order to understand the nature of these minima, in Fig.~\ref{Fig2}, the shell corrections $E_{sh}$ are shown by the red lines. The deformation dependence of $E_{sh}$ are obtained using the Strutinsky’s method \cite{Strutinsky1967,Brack1972} with the single-particle energies calculated with the Nilsson model for given quadrupole deformations $\beta_{2}$.

As expected, the $E_{sh}(\beta_{2})$ curve shows clear minima corresponding to shell effects.
Large negative values of $E_{sh}$, corresponding to shell gaps or increased level bunching near the Fermi energy, tend to reduce $E_{eq}$. This occurs because a large negative shell correction corresponds to a wide gap separating groups of opposite–parity orbitals. This observation is in line with earlier studies at spherical shell closures \cite{Mocelj2007}. For both $^{240,242}$Pu isotopes, the combined influence of deformation and shell effects leads to a pronounced reduction of $E_{eq}$ at the second minimum compared to the ground-state deformation, suggesting enhanced availability and coupling of opposite-parity configurations at large deformation.

It should be emphasized that the minimum of the total potential energy does not necessarily coincide with the minimum of the shell correction. The total energy is the sum of the macroscopic contribution and the microscopic shell correction, and the precise location of the potential–energy minimum is determined by the balance between these two terms. Since parity equilibration is directly connected with the structure of single–particle levels, in this work we focus not on the total potential energy but on the region of pronounced minima in the shell correction, as the relevant region where parity equilibration shows systematic behavior.

\section{\label{Sec4} Conclusions}
Our results show that parity equilibration occurs at lower excitation energies in the second minimum compared to the ground-state minimum, indicating that parity asymmetry in the level density persists longer in the GS configuration. This behavior reflects the combined influence of shell effects and nuclear deformation on the availability and mixing of opposite-parity single-particle states. The parity-asymmetry in the level density introduces an additional energy-dependent hindrance to the changes of shape associated with deformations of odd and even mutipolarities. Hence, these results may have implications for the statistical description of fission dynamics.


\end{document}